\documentclass[english, american]{IEEEtran}
\usepackage[T1]{fontenc}
\usepackage[latin9]{inputenc}
\usepackage{color}
\usepackage{array}
\usepackage{booktabs}
\usepackage{mathtools}
\usepackage{amsmath}
\usepackage{amssymb}
\usepackage{cancel}
\usepackage{mathdots}
\usepackage{stmaryrd}
\usepackage{stackrel}
\usepackage{graphicx}
\PassOptionsToPackage{version=3}{mhchem}
\usepackage{mhchem}
\usepackage{esint}

\makeatletter

\usepackage{colortbl}
\usepackage{cite}

\usepackage{flushend}
\usepackage{stfloats}
\usepackage[margin=8pt,font=footnotesize,labelfont=md,labelsep=colon]{caption}

\usepackage{multicol}
\usepackage{float}
\usepackage{lipsum}
\usepackage[figurename=Fig.]{caption}

\makeatother

\usepackage{babel}
\begin{document}
\title{Reconfigurable Intelligent Surfaces-Enabled Vehicular Networks: A
Physical Layer\\ Security Perspective}
\author{\textcolor{black}{\normalsize{}Abubakar Makarfi}\textit{\textcolor{black}{\normalsize{},
Member, IEEE,}}\textcolor{black}{\normalsize{} Khaled M. Rabie}\textit{\textcolor{black}{\normalsize{},
Member, IEEE}}\textcolor{black}{\normalsize{}, Omprakash Kaiwartya,
}\textit{\textcolor{black}{\normalsize{}Member, IEEE}}\textcolor{black}{\normalsize{},
Kabita Adhikari, }\textit{\textcolor{black}{\normalsize{}Member, IEEE}}\textcolor{black}{\normalsize{},
Xingwang Li}\textit{\textcolor{black}{\normalsize{}, Senior Member, IEEE}}\textcolor{black}{\normalsize{},
Marcela Quiroz-Castellanos, Rupak Kharel, }\textit{\textcolor{black}{\normalsize{}Senior
Member, IEEE}}\textcolor{black}{\normalsize{}.}\\
\foreignlanguage{american}{\textcolor{black}{}}\thanks{\textcolor{black}{A. U. Makarfi and R. Kharel are with the Department
of Computing and Mathematics, Manchester Metropolitan University,
Manchester, UK, M15 6BH (e-mails: \{a.makarfi; r.kharel\}@mmu.ac.uk).
}\protect \\
\textcolor{black}{K. M. Rabie is with the Department of Engineering,
Manchester Metropolitan University, Manchester, UK, M15 6BH (e-mail:
k.rabie@mmu.ac.uk). }\protect \\
\textcolor{black}{O. Kaiwartya is with the School of Science and Technology,
Nottingham Trent University, UK (e-mail: omprakash.kaiwartya@ntu.ac.uk).}\protect \\
\textcolor{black}{K. Adhikari is with the School of Engineering, Newcastle
University, Newcastle, UK, NE1 7RU (email: kabita.adhikari@ncl.ac.uk).}\protect \\
\textcolor{black}{X. Li is with the School of Physics and Electronic
Information Engineering, Henan Polytechnic University, Jiaozuo, China
(e-mail: lixingwang@hpu.edu.cn).}\protect \\
\textcolor{black}{M. Quiroz-Castellanos is with the Artificial Intelligence
Research Center, Universidad Veracruzana, Mexico (e-mail: maquiroz@uv.mx).}}}
\maketitle
\begin{abstract}
This paper studies the physical layer security (PLS) of a vehicular
network employing reconfigurable intelligent surfaces (RISs). RIS
technologies are emerging as an important paradigm for the realisation
of next-generation smart radio environments, where large numbers of
small, low-cost and passive elements, reflect the incident signal
with an adjustable phase shift without requiring a dedicated energy
source. Inspired by the promising potential of RIS-based transmission,
we investigate the PLS of two vehicular network system models: One
with vehicle-to-vehicle communication with the source employing a
RIS-based access point, and the other is in the form of a vehicular
adhoc network (VANET), with a RIS-based relay deployed on a building;
both models assume the presence of an eavesdropper. The performance
of the proposed systems are evaluated in terms of the average secrecy
capacity (ASC) and the secrecy outage probability (SOP). We present
accurate analytical expressions for the two performance metrics and
study the impact of various system parameters on the overall performance
of the two considered system configurations. Monte-Carlo simulations
are provided throughout to validate the results. The results show
that performance of the system in terms of the ASC and SOP is affected
by the location of the RIS-relay as well as the number of RIS cells.
Moreover, upto an order magnitude gain could be achieved within certain
regions when the number of RIS cells are doubled, clearly indicating
the benefit of employing a RIS configuration.
\end{abstract}


\section{\textcolor{black}{Introduction}}

\textcolor{black}{Reconfigurable intelligent surfaces (RISs), otherwise
known as Large Intelligent Surfaces, Intelligent Reflecting Surfaces,
reflector-arrays or intelligent walls, have recently gained much research
attention for beyond 5G technology applications. This emerging communication
paradigm is considered a potential enabling technology in realising
the concept of ``smart radio environments'' employed at the physical
layer to control }the propagation environment in order to improve
signal quality and coverage \textcolor{black}{\cite{Renzo2019,Liaskos_metasurface}.
RISs are man-made surfaces of electromagnetic material (known as a
metasurface) that are electronically controlled with integrated electronics
and have unique wireless communication capabilities \cite{Basar2019WirelessCT}.
These man-made surfaces are composed of arrays of passive scattering
elements with specially designed physical structures, where each scattering
element can be controlled in a software-defined manner to change the
phase shift and other signal characteristics, of the incident signals
on the scattering elements \cite{gong2019RISsurvey}. The possibility
of controlling the reflective interface converts the traditional random
radio environment into a smart space to support several applications
beneficial to wireless communications \cite{Basar2019WirelessCT,huangHoloMIMO,liang2019LIS}. }

RIS-based transmission concepts have been shown to be completely different
from existing multiple-input multiple-output (MIMO), beamforming and
amplify-and-forward/decode-and-forward relaying paradigms. The large
number of small, low-cost and passive elements on a RIS, only reflect
the incident signal with an adjustable phase shift without requiring
a dedicated energy source for RF processing or retransmission \cite{Basar_xmsn_LIS},
which is particularly beneficial for its energy efficiency feature.
Key benefits of the RIS-enabled communication in beyond 5G applications
have been investigated with respect to the propagation channel \cite{makarfiRISfisher,IRSphysics,IRS_space_comm},
energy efficiency \cite{EE_MU_MISO_LIS,Huang2018ReconfigurableIS},
signal-to-noise ratio (SNR) maximisation \cite{Basar_xmsn_LIS}, improving
signal coverage \cite{subrt}, improving massive MIMO systems \cite{MIMO_LIS,beyondMIMO_positioningRIS,DL_LIS_mmWave_MIMO},
improved interference suppression capability \cite{Chen_2016,HollowayReview},
beamforming optimisation \cite{Wu_beam_opt,Wu_beamform,7510962,IRS_active_passive_beam}
and multi-user networks\cite{RIS_MU_MIMO,EE_MU_MISO_LIS,LIS_multi_user},
for enhanced capacity, spectral efficiency and higher rates. Benefits
have also been demonstrated for RIS-assisted physical layer security
(PLS) \cite{secure_IRS,Chen2019IntelligentRS,enablingSecureRIS,IRSsecureChu,sec_max_IRS}
purposes, due to the flexibility of simultaneously enhancing or suppressing
signal beams to different users \cite{gong2019RISsurvey}. It is worth
noting that, PLS is achieved by employing the inherent characteristics
of the propagation channels, such as interference, fading and noise
to realise keyless secure transmission through signal processing approaches.

With regards to PLS in vehicular communications, the rapid advancements
towards autonomous vehicles and smart/cognitive transportation networks,
have made the security of vehicular networks become even more important
in order to ensure safety and minimise the risk of abuse or attacks
\cite{relay_dbl_ray,fuzzyVCPS,geometryGPSoutage_ompra,measuresVCPS}.
Although several researchers have studied PLS in vehicular networks
without the use of RISs \cite{psl_v2v,PLSinterferenceV2v,Pandey2018,plsXu2019,PandeyAFpls2018},
or non-vehicular RIS-assisted PLS for wireless applications \cite{gong2019RISsurvey},
none of these works studied PLS of RIS-enabled vehicular networks
as well as PLS of such networks. 

Inspired by the promising potential of RIS-based transmission for
both PLS and vehicular network communication, as well as the importance
of security to vehicular networks, this paper is therefore dedicated
to studying the PLS of two vehicular network models. We first consider
a vehicle-to-vehicle (V2V) network with the source employing a RIS-based
access point (for transmission), and in the second model we consider
a vehicular adhoc network (VANET), with a RIS-based relay deployed
on a building. Both models assume the presence of an eavesdropper.
The performance of the proposed systems are evaluated in terms of
the average secrecy capacity (ASC) and the secrecy outage probability
(SOP). 

There is a four-fold motivation for this research. Firstly, RIS-based
technologies can be targeted for mobile nodes in vehicular networks
due to the ease of deployability and flexible configuration that allows
for deployment in several shapes, locations and sizes, from tens to
hundreds of cells \cite{Basar2019WirelessCT}. Secondly, existing
RIS related literature have shown that most RIS-enabled applications
are designed with the RIS as a reflector \cite{Basar2019WirelessCT,gong2019RISsurvey},
since the key benefit of employing a RIS is to control the radio environment
for transmissions through smart reflections. However, other studies
have adopted the RIS as a transmitter (or access point) \cite{Basar2019WirelessCT,Basar_xmsn_LIS}
or RIS as a signal receiver \cite{Jung2018PerformanceAO}. In this
research, we consider two models adopting the RIS as a reflector and
as a transmitter as applicable to vehicular networks. Thirdly, the
mobility effect of nodes have been shown to impact the performance
of vehicular nodes (see for e.g., \cite{khattabi_MobileNode}), although
several previous studies assume quasi-static node positions due to
ease of analysis. It has however been shown recently, that rapid fluctuations
in the received signal strength due to the doppler effect can be effectively
reduced by using the real-time tuneable RISs\footnote{This study will not go into detail on the process of mitigating doppler
effects with RISs. The reader is however referred to \cite{basar2019dopplerRIS}
for a more detailed analysis.} \cite{basar2019dopplerRIS}. Fourthly, the need for accurate analytical
analysis for easily determining PLS parameters given that the existing
literature on RIS-assisted PLS for non-vehicular applications have
only reported some initial simulation results \cite{gong2019RISsurvey}.

\textcolor{black}{From the aforementioned, this study therefore presents
the following contributions:}
\begin{itemize}
\item We present a novel analysis of possible implementation of RIS to vehicular
networks. To the best of the authors' knowledge, this is the first
analysis of RIS-enabled vehicular networks, as well as the first time
study of the PLS of such networks.
\item \textcolor{black}{We study the PLS of two possible vehicular network
models. The first model considers a V2V network with the source vehicle
employing a RIS-based access point (AP) for transmission. In the second
model, we consider a VANET with a source station transmission via
a RIS-based relay. Such a RIS-relay could be deployed on a building
as part of a smart infrastructure within a smart city environment. }
\item The PLS analysis \textcolor{black}{is achieved by analyzing and deriving
expressions for the ASC and the SOP of the system. The derived expressions
allow for the ease of investigation of key system parameters.}
\item For the ASC, accurate closed-form approximations are further presented
to make the analysis of key parameters more tractable. On comparison
with the exact derived expressions, it will be observed that these
approximations are highly accurate within the regions of interest.
Moreover, with the approximate expressions, it becomes easier to gain
insights into the behavior of the considered scenarios.
\item \textcolor{black}{The distances of the legitimate and eavesdropper
nodes are taken into account along with realistic fading scenarios
considered for the base stations and the vehicular mobile nodes. }
\end{itemize}
\textcolor{black}{For all the scenarios considered in this work, Monte
Carlo simulations are provided to verify the accuracy of the analysis.
The results show that the performance of the system in terms of the
secrecy capacity is improved with the use of the RIS. Furthermore,
the effect of the system parameters such as source power, eavesdropper
distance and number of RIS cells on the system performance are investigated.
Mathematical functions and notations are presented in Table I.}

\textcolor{black}{The paper is organized as follows. In Sections \ref{sec:RIS AP}
and \ref{sec:RIS-relay}, we describe the two system models under
study and analyze the secrecy performance by deriving accurate analytical
expressions for efficient computation of the secrecy capacity of the
networks. Thereafter, in Sections \ref{sec:Results} and \ref{sec:Conclusions},
we present the results with discussions and outline the main conclusions,
respectively.}

\subsection*{Mathematical Functions and Notations}

The following notations are used: \\

\begin{tabular}{l>{\raggedright}p{5.5cm}}
\toprule 
Notation & Definition\tabularnewline
\midrule
\midrule 
$\mathbb{E}\left[\cdot\right]$ & Expectation operator\tabularnewline
\midrule 
\textcolor{black}{$\textrm{G}_{u,v}^{s,t}\left(x\mid\cdots\right)$} & \textcolor{black}{Meijer\textquoteright s G-function \cite[Eq. (9.302)]{bookV3}}\tabularnewline
\midrule 
$\Gamma\left(z\right)$ & $=\int_{0}^{\infty}t^{z-1}e^{-t}dt$, the gamma function \cite[Eq. (8.310)]{book2}\tabularnewline
\midrule 
$\textrm{erf}\left(x\right)$ & =$\frac{2}{\sqrt{\pi}}\int_{0}^{x}e^{-t^{2}}dt$, the error function
\cite[Eq. (8.250.1)]{book2}\tabularnewline
\midrule 
$_{\phantom{}2}F_{1}\left(\alpha;\beta;\gamma;z\right)$ & Gauss hypergeometric function \cite[Eq. (9.111)]{book2}\tabularnewline
\midrule 
$K_{v}\left(z\right)$ & Modified Bessel function of the second kind and $v$th order \cite[Eq. (8.407)]{book2}\tabularnewline
\midrule 
$\mathcal{M}_{X}\left(z\right)$ & $=\mathbb{E}\left[e^{-zX}\right]$, the moment generating function
of $X$.\tabularnewline
\midrule 
$f\left(\gamma\right)$ & Probability density function of $\gamma$.\tabularnewline
\bottomrule
\end{tabular}

\section{V2V with RIS as Access Point\label{sec:RIS AP}}

In this section, we describe the V2V network with RIS as AP and derive
expressions for the ASC and SOP of the system.

\subsection{System Description}

\begin{figure}[th]
\begin{centering}
\includegraphics[scale=0.4]{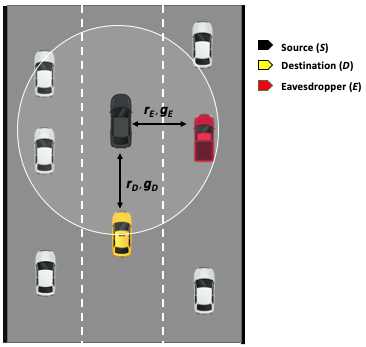}
\par\end{centering}
\caption{System model for V2V scenario with the source vehicle using RIS as
AP.\label{fig:sys-mod}}
\end{figure}

\begin{figure}[th]
\begin{centering}
\includegraphics[scale=0.35]{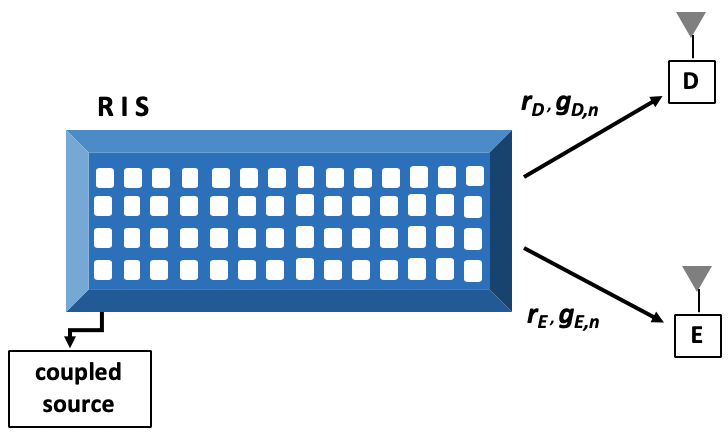}
\par\end{centering}
\caption{Vehicle RIS configuration as an AP.\label{fig:sys-mod-2}}
\end{figure}

We consider a system of vehicles operating in a network as shown in
Fig. \ref{fig:sys-mod}. We assume a classic Wyner's wiretap model
in our analysis \cite{Lei_sec_cap}, such that an information source
vehicle ($S$), sends confidential information to a destination vehicle
($D$), while a passive eavesdropper vehicle ($E$) attempts to receive
and decode the confidential information. The vehicles $D$ and $E$
are known to lie within a certain radius from $S,$ the precise relative
distances of the V2V links are unknown during transmission, which
is a realistic assumption for a network of this nature \cite{pls_eve_uncertain,psl_v2v}.
Moreover, $S$ is assumed to employ a RIS-based scheme in the form
of an AP to communicate over the network\footnote{Initial proposal and results for such a configuration of intelligent
surfaces employed as a transmitter or access point was reported in
\cite{Basar_xmsn_LIS}.}. As shown in the block diagram of Fig. \ref{fig:sys-mod-2}, the
RIS can be connected over a wired link or optical fiber for direct
transmission from $S$, and can support transmission without RF processing.
For the system considered, we assume an intelligent AP with the RIS
having knowledge of channel phase terms, such that the RIS-induced
phases can be adjusted to maximize the received SNR through appropriate
phase cancellations and proper alignment of reflected signals from
the intelligent surface.

The received signals at $D$ and $E$ are respectively represented
as
\begin{align}
y_{D} & =\left[\sum_{n=1}^{N}h_{D,n}e^{-j\phi_{n}}\right]x+w_{D},\label{eq:yD}\\
y_{E} & =\left[\sum_{n=1}^{N}h_{E,n}e^{-j\phi_{n}}\right]x+w_{E},\label{eq:yE}
\end{align}
where $x$ represents the transmitted signal by $S$ with power $P_{s}$,
while the terms $w_{D}$ and $w_{E}$ are the respective additive
white Gaussian noise (AWGN) at $D$ and $E$, respectively. Without
loss of generality, we denote the power spectral density of the AWGN
as $N_{0}$ and equal at both links. The terms $h_{i,n}=\sqrt{g_{i,n}r_{i}^{-\beta},}\quad i\in\left\{ D,E\right\} $,
is the channel coefficient from $S$ to the receiving vehicles $D$
and $E,$ where $r_{i}$ is the V2V link distance, $\beta$ is the
path-loss exponent and $g_{i,n}$ is the channel gain from the RIS
to the receiver, following independent double Rayleigh fading \cite{psl_v2v}.
The term $\phi_{n}$ is the reconfigurable phase induced by the $n$th
reflector of the RIS, which through phase matching, the SNR of the
received signals can be maximized\footnote{For the sake of brevity, the reader is referred to \cite{Basar_xmsn_LIS},
for details of phase cancellation mathematical techniques.}. It is worth noting that the electromagnetic signal reconfiguration
capability of the RIS is achieved by a joint phase control of individual
RIS cells. It is possible to implement a type of software-defined
control mechanism as proposed in \cite{Liaskos_RISdesign}, where
tunable chips are integrated within the RIS cells, such that each
chip communicates to a central controller \cite{Liu2018PM,Yang2016APM},
or the RIS 
 could be fitted with sensors to receive for intelligent control,
in-line with environmental factors \cite{Renzo2019}. 

Based on (\ref{eq:yD}) and (\ref{eq:yE}), the instantaneous SNRs
at $D$ and $E$ are given by
\begin{equation}
\gamma_{D}=\frac{\sum_{n=1}^{N}P_{s}\mid h_{D,n}\mid^{2}}{N_{0}},\label{eq:sinr-d}
\end{equation}
and

\begin{equation}
\gamma_{E}=\frac{\sum_{n=1}^{N}P_{s}\mid h_{E,n}\mid^{2}}{N_{0}}.\label{eq:sinr-e}
\end{equation}

\subsection{Average Secrecy Capacity \label{sec:ASC-v2v}}

In this section, we derive analytical expressions for the ASC of the
system. The maximum achievable secrecy capacity is defined by \cite{bloch_cap}
\begin{equation}
C_{s}=\max\left\{ C_{D}-C_{E},0\right\} ,\label{eq:cs-defined}
\end{equation}
where $C_{D}=\log_{2}\left(1+\gamma_{D}\right)$ and $C_{E}=\log_{2}\left(1+\gamma_{E}\right)$
are the instantaneous capacities of the main and eavesdropping links,
respectively. The secrecy capacity in (\ref{eq:cs-defined}) can therefore
be expressed as \cite{bloch_cap}
\begin{equation}
C_{s}=\begin{cases}
\log_{2}\left(1+\gamma_{D}\right)-\log_{2}\left(1+\gamma_{E}\right), & \gamma_{D}>\gamma_{E},\\
0, & \gamma_{D}<\gamma_{E}.
\end{cases}\label{eq:cs-defined-2}
\end{equation}

The ASC $\overline{C_{s}}$ is given by \cite{Osamah18GlobecomSC}
\begin{align}
\overline{C_{s}} & =\mathbb{E}\left[C_{s}\left(\gamma_{D},\gamma_{E}\right)\right]\nonumber \\
 & =\stackrel[0]{\infty}{\int}\stackrel[0]{\infty}{\int}C_{s}\left(\gamma_{D},\gamma_{E}\right)f\left(\gamma_{D},\gamma_{E}\right)\textrm{d}\gamma_{D}\textrm{d}\gamma_{E},\label{eq:av-sec-cap-1}
\end{align}
where $\mathbb{E}\left[\cdot\right]$ is the expectation operator
and $f\left(\gamma_{D},\gamma_{E}\right)$ is the joint PDF of $\gamma_{D}$
and $\gamma_{E}$. In order to simplify the analysis, we express the
logarithmic function in (\ref{eq:cs-defined}) in an alternate form.
Recalling the identity \cite[Eq. (6)]{hamdi_cap_mrc}
\begin{equation}
\textrm{ln}\left(1+\zeta\right)=\stackrel[0]{\infty}{\int}\frac{1}{s}\left(1-e^{-\zeta s}\right)e^{-s}\textrm{d}s,\label{eq:log-id-1}
\end{equation}
and by substituting $\zeta=\gamma_{D}$ in (\ref{eq:log-id-1}), we
can express the instantaneous capacity of the main link as
\begin{equation}
\overline{C}_{D}=\frac{1}{\textrm{ln}\left(2\right)}\stackrel[0]{\infty}{\int}\frac{1}{z}\left(1-\mathcal{M}_{D}\left(z\right)\right)e^{-z}\textrm{d}z,\label{eq: log-id-2}
\end{equation}
where $\mathcal{M}_{D}\left(z\right)=\mathbb{E}\left[e^{-z\frac{P_{s}r_{D}^{-\beta}}{N_{0}}\sum_{n=1}^{N}g_{D,n}}\right]$
is the moment generating function (MGF) of the SNR at $D$. 

Next, we compute the MGF $\mathcal{M}_{D}\left(z\right)$, defined
by
\begin{align}
\mathcal{M}_{D}\left(z\right)= & \mathbb{E}\left[e^{-z\frac{P_{s}r_{D}^{-\beta}}{N_{0}}\sum_{n=1}^{N}g_{D,n}}\right]\nonumber \\
= & \prod_{n=1}^{N}\mathbb{E}\left[e^{-z\frac{P_{s}r_{D}^{-\beta}}{N_{0}}g_{D,n}}\right]\nonumber \\
= & \prod_{n=1}^{N}\stackrel[g_{D}]{}{\int}e^{-z\xi_{D}g_{D,n}}f_{g_{D}}(g)\textrm{d}g_{D},\label{eq:mgf-D-1}
\end{align}
then from the generalized cascaded Rayleigh distribution, we can obtain
the PDF of the double Rayleigh channel for $n=2$ in \cite[Eq. (8)]{cascaded_ray}
as $f\left(g\right)=\textrm{G}_{0,2}^{2,0}\left(\frac{1}{4}g^{2}\Biggl|\negthickspace\begin{array}{c}
-\\
\frac{1}{2},\negthickspace\frac{1}{2}
\end{array}\negthickspace\right).$ By invoking \cite[Eq. (9.34.3)]{book2}, we can express the PDF by
re-writing the Meijer G-function in an alternate form. Thus, we get
\begin{equation}
f\left(g\right)=\textrm{G}_{0,2}^{2,0}\left(\frac{1}{4}g^{2}\Biggl|\negthickspace\begin{array}{c}
-\\
\frac{1}{2},\negthickspace\frac{1}{2}
\end{array}\negthickspace\right)=gK_{0}\left(g\right).\label{eq:meijer-bessel}
\end{equation}

Using (\ref{eq:meijer-bessel}) and \cite[Eq. (6.621.3)]{book2} along
with some basic algebraic manipulations, we can obtain the desired
result as 
\begin{equation}
\mathcal{M}_{D}\left(z\right)=\prod_{n=1}^{N}\frac{4}{3(1+z\frac{P_{s}r_{D}^{-\beta}}{N_{0}})^{2}}{}_{\phantom{}2}F_{1}\left(2,\frac{1}{2},\frac{5}{2},\frac{z\frac{P_{s}r_{D}^{-\beta}}{N_{0}}-1}{z\frac{P_{s}r_{D}^{-\beta}}{N_{0}}+1}\right).\label{eq:mgf-D-final}
\end{equation}

\textcolor{black}{Using similar analysis, the average capacity of
the eavesdropper link can be represented as}
\begin{equation}
\overline{C}_{E}=\frac{1}{\textrm{ln}\left(2\right)}\stackrel[0]{\infty}{\int}\frac{1}{z}\left(1-\mathcal{M}_{E}\left(z\right)\right)e^{-z}\textrm{d}z,\label{eq:av-CE-defined}
\end{equation}
where the MGF $\mathcal{M}_{E}\left(z\right)=\mathbb{E}\left[e^{-z\frac{P_{s}r_{E}^{-\beta}}{N_{0}}\sum_{n=1}^{N}g_{E}}\right]$
and can be similarly evaluated as 
\begin{equation}
\mathcal{M}_{E}\left(z\right)=\prod_{n=1}^{N}\frac{4}{3(1+z\frac{P_{s}r_{E}^{-\beta}}{N_{0}})^{2}}{}_{\phantom{}2}F_{1}\left(2,\frac{1}{2},\frac{5}{2},\frac{z\frac{P_{s}r_{E}^{-\beta}}{N_{0}}-1}{z\frac{P_{s}r_{E}^{-\beta}}{N_{0}}+1}\right).\label{eq:mgf-E-final}
\end{equation}

From (\ref{eq:cs-defined-2}), (\ref{eq: log-id-2}), (\ref{eq:mgf-D-final})
- (\ref{eq:mgf-E-final}), the ASC can be represented as (\ref{eq:Cs-final}),
shown at the top of the next page.

\begin{figure*}[t]
{\small{}
\begin{multline}
\overline{C}_{s}=\frac{1}{\textrm{ln}\left(2\right)}\stackrel[0]{\infty}{\int}\frac{1}{z}e^{-z}\left[\left(1-\left\{ \frac{4}{3\left(N_{0}+zP_{s}r_{D}^{-\beta}\right)^{2}}{}_{\phantom{}2}F_{1}\left(2,\frac{1}{2},\frac{5}{2},\frac{zP_{s}r_{D}^{-\beta}-N_{0}}{zP_{s}r_{D}^{-\beta}+N_{0}}\right)\right\} ^{N}\right)\right.\\
\qquad-\left.\left(1-\left\{ \frac{4}{3\left(N_{0}+zP_{s}r_{E}^{-\beta}\right)^{2}}{}_{\phantom{}2}F_{1}\left(2,\frac{1}{2},\frac{5}{2},\frac{zP_{s}r_{E}^{-\beta}-N_{0}}{zP_{s}r_{E}^{-\beta}+N_{0}}\right)\right\} ^{N}\right)\right]\textrm{d}z.\label{eq:Cs-final}
\end{multline}
}{\small\par}
\selectlanguage{american}%
\centering{}\rule[0.5ex]{2.03\columnwidth}{0.8pt}\selectlanguage{american}%
\end{figure*}

\subsection{Approximate Secrecy Capacity\label{subsec:Aprx-sec-cap-model1}}

In this section, we present an approximate solution to the ASC computed
in Sec. \ref{sec:ASC-v2v}, in order to provide a more direct solution.
From (\ref{eq:cs-defined-2}), it can be seen that the secrecy capacity
is defined as a logarithmic function. Thus, we can define an approximate
bound to the solution in (\ref{eq:Cs-final}) by invoking Jensen's
inequality\footnote{Jensen's inequality asserts that, if $f\left(x\right)$ is a convex
function, then $\mathbb{E}\left[f\left(X\right)\right]\geq f\left(\mathbb{E}\left[X\right]\right)$,
provided that the expectations exist and are finite.} \cite[pp. 453]{Ross_probability} and applying to the expressions
for the instantaneous capacities of the main and eavesdropping links.
Therefore, the average capacity at $D$ is 
\begin{align}
\mathbb{E}\left[\log_{2}\left(1+\gamma_{D}\right)\right] & \leq\log_{2}\left(1+\mathbb{E}\left[\gamma_{D}\right]\right)\nonumber \\
 & =\log_{2}\left(1+\mathbb{E}\left[\frac{P_{s}r_{D}^{-\beta}}{N_{0}}\sum_{n=1}^{N}g_{D,n}\right]\right)\nonumber \\
 & =\log_{2}\left(1+\frac{P_{s}r_{D}^{-\beta}}{N_{0}}\mathbb{E}\left[\sum_{n=1}^{N}g_{D,n}\right]\right),\label{eq:cd-aprx-1}
\end{align}
where $\gamma_{D}$ is defined in (\ref{eq:sinr-d}). By using (\ref{eq:meijer-bessel}),
the expectation in (\ref{eq:cd-aprx-1}) can be obtained as 
\begin{align}
\mathbb{E}\left[\sum_{n=1}^{N}g_{D,n}\right] & =\stackrel[0]{\infty}{\int}\sum_{n=1}^{N}g_{D,n}g_{D}K_{0}\left(g_{D}\right)\textrm{d}g_{D}\nonumber \\
 & \stackrel{\left(a\right)}{=}N\frac{\pi}{2},\label{eq:mean-bessel}
\end{align}
where $(a)$ in (\ref{eq:mean-bessel}) was obtained with the aid
of \cite[Eq. (6.521.10)]{book2}. From (\ref{eq:cd-aprx-1}) and (\ref{eq:mean-bessel}),
we obtain a bounded expression for the average capacity at $D$ as
\begin{equation}
\overline{C}_{D}^{aprx}=\log_{2}\left(1+\frac{N\pi P_{s}r_{D}^{-\beta}}{2N_{0}}\right).\label{eq:cd-aprx-final}
\end{equation}

Using similar analysis to the derivation of (\ref{eq:cd-aprx-final}),
we obtain a bounded expression for the average capacity at $E$ as
\begin{equation}
\overline{C}_{E}^{aprx}=\log_{2}\left(1+\frac{N\pi P_{s}r_{E}^{-\beta}}{2N_{0}}\right).\label{eq:ce-aprx-final}
\end{equation}

From (\ref{eq:cs-defined-2}), (\ref{eq:cd-aprx-final}) and (\ref{eq:ce-aprx-final}),
we obtained the desired bounded expression as 
\begin{equation}
\overline{C}_{s}^{aprx}=\log_{2}\left(\frac{2N_{0}+N\pi P_{s}r_{D}^{-\beta}}{2N_{0}+N\pi P_{s}r_{E}^{-\beta}}\right).\label{eq:cs-aprx-final}
\end{equation}

It is worth noting that, in addition to the fact that (\ref{eq:cs-aprx-final})
lends itself much more easily to analysis as compared to (\ref{eq:Cs-final}),
the expression (\ref{eq:cs-aprx-final}) also produces highly accurate
results for the parameters of interest, as will be demonstrated in
the discussion in Sec. \ref{sec:Results}.

\subsection{Secrecy Outage Probability\label{subsec:SOP-1}}

In this section, we derive an expression for the SOP of the V2V RIS
AP model. The SOP is defined as the probability that the secrecy capacity
falls below a target secrecy rate \cite{PLSinterferenceV2v}. This
can be represented as
\begin{equation}
P_{o}=\textrm{Pr}\left[C_{s}<c_{\textrm{th}}\right]\label{eq:pout-defined}
\end{equation}
where $c_{\textrm{th}}$ is the pre-determined target secrecy rate.
From (\ref{eq:cs-defined-2}) and (\ref{eq:pout-defined}) we obtain

\begin{align}
P_{o} & =\textrm{Pr}\left[\log_{2}\left(\frac{1+\gamma_{D}}{1+\gamma_{E}}\right)<c_{\textrm{th}}\right]\nonumber \\
 & =\textrm{Pr}\left[\frac{1+\gamma_{D}}{1+\gamma_{E}}<2^{c_{\textrm{th}}}\right]\nonumber \\
 & \stackrel{\left(b\right)}{=}\textrm{Pr}\left[P_{s}r_{D}^{-\beta}\sum_{n=1}^{N}g_{D,n}<\nu\left(N_{0}+P_{s}r_{E}^{-\beta}\sum_{n=1}^{N}g_{E,n}\right)-N_{0}\right]\nonumber \\
 & =\textrm{Pr}\left[\sum_{n=1}^{N}g_{D,n}<\frac{\nu\left(N_{0}+P_{s}r_{E}^{-\beta}\sum_{n=1}^{N}g_{E,n}\right)-N_{0}}{P_{s}r_{D}^{-\beta}}\right]\nonumber \\
 & \stackrel{\left(c\right)}{=}\textrm{Pr}\left[\sum_{n=1}^{N}g_{D,n}<\underset{\varTheta}{\underbrace{\frac{N_{0}\left(\nu-1\right)}{P_{s}r_{D}^{-\beta}}+\frac{\nu r_{E}^{-\beta}}{r_{D}^{-\beta}}\sum_{n=1}^{N}g_{E,n}}},\right]\label{eq:pout-defined-2}
\end{align}
where $\left(b\right)$ follows from substituting (\ref{eq:sinr-d})
and (\ref{eq:sinr-e}) and $\nu=2^{c_{\textrm{th}}}$. Given the difficulty
in obtaining a tractable expression for the distribution of the sum
$\sum_{n=1}^{N}g_{D,n}$ in $\left(c\right)$ of (\ref{eq:pout-defined-2}),
we employ an appropriate approximation using the central limit theorem
(CLT) for reasonably large number of reflecting cells when $N\gg1$.
It therefore follows that $\sum_{n=1}^{N}g_{D,n}$ can be approximated
by a Gaussian distribution with parameters; mean $\mu=N\frac{\pi}{2}$
(from (\ref{eq:mean-bessel})) and variance $\sigma^{2}$ given by
\begin{align}
\textrm{Var}\left[\sum_{n=1}^{N}g_{D,n}\right] & =\mathbb{E}\left[\sum_{n=1}^{N}g_{D,n}^{2}\right]-\mathbb{E}\left[\sum_{n=1}^{N}g_{D,n}\right]^{2}\nonumber \\
 & =N\left(4-\frac{\pi^{2}}{4}\right).\label{eq:var-of-sum}
\end{align}

Thus from (\ref{eq:pout-defined-2}), the SOP can be given by 
\begin{equation}
P_{o}\approx\frac{1}{2}\left(1+\textrm{erf}\left(\frac{\varTheta-\mu}{\sqrt{2}\sigma}\right)\right),\label{eq:sop-aprx-defined}
\end{equation}
where $\varTheta$ is defined in (\ref{eq:pout-defined-2}) and $\mathbb{E}\left[\varTheta\right]$
can be computed as
\begin{align}
\mathbb{E}\left[\varTheta\right] & =\mathbb{E}\left[\frac{N_{0}\left(\nu-1\right)}{P_{s}r_{D}^{-\beta}}+\frac{\nu r_{E}^{-\beta}}{r_{D}^{-\beta}}\sum_{n=1}^{N}g_{E,n}\right]\nonumber \\
 & =\mathbb{E}\left[\frac{N_{0}\left(\nu-1\right)}{P_{s}r_{D}^{-\beta}}+\frac{\nu r_{E}^{-\beta}}{r_{D}^{-\beta}}\sum_{n=1}^{N}g_{E,n}\right]\nonumber \\
 & \stackrel{\left(d\right)}{=}\frac{N_{0}\left(\nu-1\right)}{P_{s}r_{D}^{-\beta}}+\frac{\nu N\pi r_{E}^{-\beta}}{2r_{D}^{-\beta}},\label{eq:mean-theta}
\end{align}
\begin{figure}[th]
\begin{centering}
\includegraphics[scale=0.45]{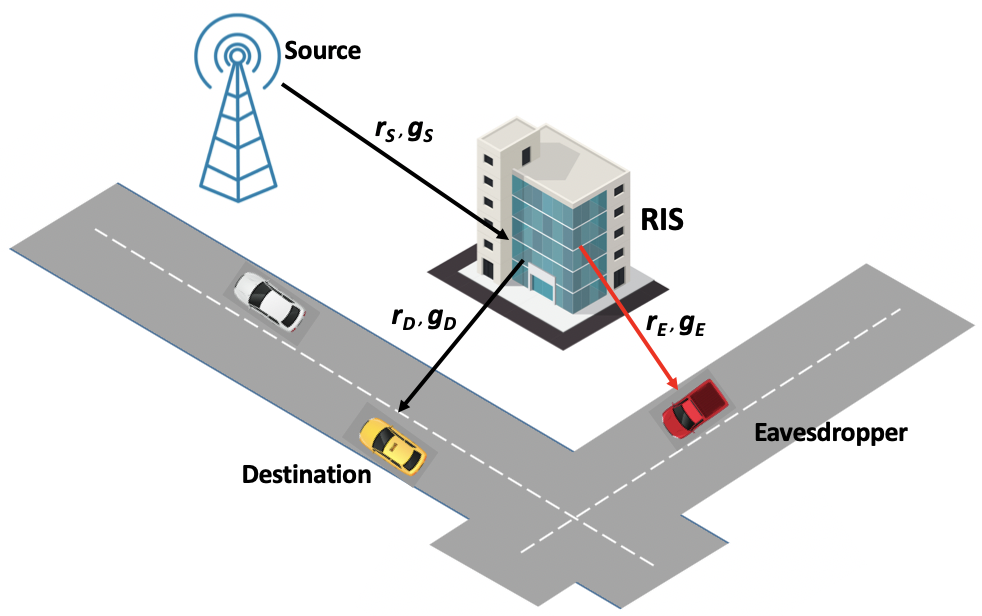}
\par\end{centering}
\caption{System model for a vehicular network scenario. Source station using
building-mounted-RIS as relay for vehicular communication.\label{fig:sys-mod-1}}
\end{figure}
where $\left(d\right)$ in (\ref{eq:mean-theta}) follows from (\ref{eq:mean-bessel})
with $\mathbb{E}\left[\sum_{n=1}^{N}g_{E,n}\right]=N\frac{\pi}{2}.$
The SOP for the V2V RIS access point is thus
\begin{equation}
P_{o}\approx\frac{1}{2}\left(1+\textrm{erf}\left(\frac{\frac{N_{0}\left(\nu-1\right)}{P_{s}r_{D}^{-\beta}}+N\frac{\pi}{2}\left(\frac{\nu r_{E}^{-\beta}}{r_{D}^{-\beta}}-1\right)}{\sqrt{2N\left(4-\frac{\pi^{2}}{4}\right)}}\right)\right).\label{eq:sop-aprx-final}
\end{equation}

\section{VANET Transmission Through RIS Relay\label{sec:RIS-relay} }

In this section, we describe the VANET system transmitting through
a RIS relay and derive expressions for the ASC and SOP of the system.

\subsection{System Description}

In this section, we consider an RIS-based scheme with the RIS employed
as a relay or reflector for vehicular nodes in the network. Fig. \ref{fig:sys-mod-1}
illustrates the RIS-based system under consideration. The RIS is deployed
on a building and used as a relay for the signal from stationary source
$S$, while $D$ and $E$ are assumed to be highly mobile vehicular
nodes. Under this assumption, the source-to-RIS channel $g_{s}$ is
assumed to be Rayleigh faded, while the RIS-to-destination and RIS-to-eavesdropper
fading channels, $g_{D}$ and $g_{E}$ are assumed to be double-Rayleigh
distributed. The RIS is in the form of a reflect-array comprising
N reconfigurable reflector elements, capable of being controlled by
a communication oriented software for intelligent transmission. With
this in mind, the received signals at $D$ and $E$ are
\begin{equation}
y_{i}=\left[\sum_{n=1}^{N}h_{s}h_{i,n}e^{-j\phi_{n}}\right]x+w_{i},i\in\{D,E\},\label{eq:y}
\end{equation}
where $h_{s,n}=\sqrt{g_{s,n}r_{s}^{-\beta}e^{-j\theta_{n}}}$ is the
source-to-RIS channel with distance $r_{s}$, phase component $\theta_{n}$
and $g_{s}$ following a Rayleigh fading distribution. The term $h_{i,n}=\sqrt{g_{i,n}r_{i}^{-\beta}e^{-j\psi_{n}}},\quad i\in\left\{ D,E\right\} $,
is the channel coefficient from RIS-to-vehicle node, with distance
$r_{i}$, path-loss exponent $\beta$, phase component $\psi_{n}$
and $g_{i,n}$ following a double-Rayleigh distribution to model the
mobility of the nodes \cite{psl_v2v}. The instantaneous SNRs at $D$
and $E$ are given respectively by
\begin{equation}
\gamma_{D}^{r}=\frac{\sum_{n=1}^{N}P_{s}\mid h_{s,n}\mid^{2}\mid h_{D,n}\mid^{2}}{N_{0}}\label{eq:snr-D-r}
\end{equation}
and
\begin{equation}
\gamma_{E}^{r}=\frac{\sum_{n=1}^{N}P_{s}\mid h_{s,n}\mid^{2}\mid h_{E,n}\mid^{2}}{N_{0}}\label{eq:snr-E-r}
\end{equation}

\subsection{Average Secrecy Capacity Analysis\label{subsec:ASC-VANET}}

From (\ref{eq: log-id-2}), we obtain the average capacity for the
destination V2V link as 
\begin{equation}
\overline{C}_{D,r}=\frac{1}{\textrm{ln}\left(2\right)}\stackrel[0]{\infty}{\int}\frac{1}{z}\left(1-\mathcal{M}_{D,r}\left(z\right)\right)e^{-z}\textrm{d}z,\label{eq: log-id-3}
\end{equation}
where $\mathcal{M}_{D,r}\left(z\right)=\mathbb{E}\left[e^{-z\frac{P_{s}r_{s}^{-\beta}r_{D}^{-\beta}}{N_{0}}\sum_{n=1}^{N}g_{s,n}g_{D,n}}\right]$
is the MGF of the SNR at $D$. As for the joint distribution of $g_{s}$
and $g_{D}$, given that $g_{s}$ is a Rayleigh RV and $g_{D}$ is
a double-Rayleigh RV, we can define the RV $g=g_{s}g_{D}$, which
follows the cascaded Rayleigh distribution with $n=3$. From the generalized
cascaded Rayleigh distribution \cite{cascaded_ray}, we can evaluate
the PDF of $g$ as

\begin{equation}
f\left(g\right)=\frac{1}{\sqrt{2}}\textrm{G}_{0,3}^{3,0}\left(\frac{1}{8}g^{2}\Biggl|\negthickspace\begin{array}{c}
-\\
\frac{1}{2},\negthickspace\frac{1}{2},\negthickspace\frac{1}{2}
\end{array}\right).\label{eq: triple rayleigh pdf}
\end{equation}

Thus, the $\mathcal{M}_{D,r}\left(z\right)$ is given by
\begin{align}
\mathcal{M}_{D,r}\left(z\right)= & \mathbb{E}\left[e^{-z\frac{P_{s}r_{s}^{-\beta}r_{D}^{-\beta}}{N_{0}}\sum_{n=1}^{N}g_{s,n}g_{D,n}}\right]\nonumber \\
= & \prod_{n=1}^{N}\stackrel[0]{\infty}{\int}e^{-zg\frac{P_{s}r_{s}^{-\beta}r_{D}^{-\beta}}{N_{0}}}f_{g}(g)\textrm{d}g\nonumber \\
= & \prod_{n=1}^{N}\stackrel[0]{\infty}{\int}\frac{1}{\sqrt{2}}e^{-zg\frac{P_{s}r_{s}^{-\beta}r_{D}^{-\beta}}{N_{0}}}\textrm{G}_{0,3}^{3,0}\left(\frac{1}{8}g^{2}\Biggl|\negthickspace\begin{array}{c}
-\\
\frac{1}{2},\negthickspace\frac{1}{2},\negthickspace\frac{1}{2}
\end{array}\right)\textrm{d}g.\label{eq:mgf-D-1-1}
\end{align}

Using (\ref{eq: triple rayleigh pdf}) and \cite[Eq. (7.813.2)]{book2}
along with some basic algebraic manipulations, we can obtain the MGF
as 
\begin{equation}
\mathcal{M}_{D,r}\left(z\right)=\prod_{n=1}^{N}\frac{1}{z\mu_{d}\sqrt{2\pi}}\textrm{G}_{2,3}^{3,2}\left(\frac{1}{2\left(z\mu_{d}\right)^{2}}\Biggl|\negthickspace\begin{array}{c}
0,\frac{1}{2}\\
\frac{1}{2},\negthickspace\frac{1}{2},\negthickspace\frac{1}{2}
\end{array}\right),\label{eq:mgf-D-final-r}
\end{equation}
where $\mu_{d}=\frac{P_{s}r_{s}^{-\beta}r_{D}^{-\beta}}{N_{0}}$\textcolor{black}{.
Using similar analysis, the average capacity of the eavesdropper link
can be represented as}
\begin{equation}
\overline{C}_{E,r}=\frac{1}{\textrm{ln}\left(2\right)}\stackrel[0]{\infty}{\int}\frac{1}{z}\left(1-\mathcal{M}_{E,r}\left(z\right)\right)e^{-z}\textrm{d}z,\label{eq:av-CE-defined-r}
\end{equation}
where the MGF $\mathcal{M}_{E,r}\left(z\right)=\mathbb{E}\left[e^{-z\frac{P_{s}r_{s}^{-\beta}r_{E}^{-\beta}}{N_{0}}\sum_{n=1}^{N}g_{s,n}g_{E,n}}\right]$
and can be similarly evaluated as 
\begin{equation}
\mathcal{M}_{E,r}\left(z\right)=\prod_{n=1}^{N}\frac{1}{z\mu_{e}\sqrt{2\pi}}\textrm{G}_{2,3}^{3,2}\left(\frac{1}{2\left(z\mu_{e}\right)^{2}}\Biggl|\negthickspace\begin{array}{c}
0,\frac{1}{2}\\
\frac{1}{2},\negthickspace\frac{1}{2},\negthickspace\frac{1}{2}
\end{array}\right),\label{eq:mgf-E-final-1}
\end{equation}
where $\mu_{e}=\frac{P_{s}r_{s}^{-\beta}r_{e}^{-\beta}}{N_{0}}$. 

From (\ref{eq:cs-defined-2}), (\ref{eq: log-id-3}), (\ref{eq:mgf-D-final-r})
- (\ref{eq:mgf-E-final-1}), the ASC can be represented as (\ref{eq:Cs-final-r}),
shown at the top of the next page.

\begin{figure*}[t]
{\small{}
\begin{multline}
\overline{C}_{s,r}=\frac{1}{\textrm{ln}\left(2\right)}\stackrel[0]{\infty}{\int}\frac{1}{z}e^{-z}\left[\left(1-\left\{ \frac{N_{0}}{\sqrt{2\pi}zP_{s}r_{s}^{-\beta}r_{D}^{-\beta}}\textrm{G}_{2,3}^{3,2}\left(\frac{1}{2}\left(\frac{N_{0}}{zP_{s}r_{s}^{-\beta}r_{D}^{-\beta}}\right)^{2}\Biggl|\negthickspace\begin{array}{c}
0,\frac{1}{2}\\
\frac{1}{2},\negthickspace\frac{1}{2},\negthickspace\frac{1}{2}
\end{array}\right)\right\} ^{N}\right)\right.\\
\qquad-\left.\left(1-\left\{ \frac{N_{0}}{zP_{s}r_{s}^{-\beta}r_{E}^{-\beta}\sqrt{2\pi}}\textrm{G}_{2,3}^{3,2}\left(\frac{1}{2}\left(\frac{N_{0}}{zP_{s}r_{s}^{-\beta}r_{E}^{-\beta}}\right)^{2}\Biggl|\negthickspace\begin{array}{c}
0,\frac{1}{2}\\
\frac{1}{2},\negthickspace\frac{1}{2},\negthickspace\frac{1}{2}
\end{array}\right)\right\} ^{N}\right)\right]\textrm{d}z.\label{eq:Cs-final-r}
\end{multline}
}{\small\par}
\selectlanguage{american}%
\centering{}\rule[0.5ex]{2.03\columnwidth}{0.8pt}\selectlanguage{american}%
\end{figure*}

\subsection{Approximate Secrecy Capacity}

In this section, we present an approximate expression to the ASC for
the VANET transmission through RIS relay, as computed in Sec. \ref{subsec:ASC-VANET}.
Following similar analysis to the derivation of an approximate expression
in Sec. \ref{subsec:Aprx-sec-cap-model1}, we invoke Jensen's inequality
to obtain a bound for the expression in (\ref{eq:Cs-final-r}). We
commence with the average capacity at $D$, which can be approximated
as 
\begin{align}
\mathbb{E}\left[\log_{2}\left(1+\gamma_{D}^{r}\right)\right] & \leq\log_{2}\left(1+\mathbb{E}\left[\gamma_{D}^{r}\right]\right)\nonumber \\
 & =\log_{2}\left(1+\mathbb{E}\left[\frac{P_{s}r_{s}^{-\beta}r_{D}^{-\beta}}{N_{0}}\sum_{n=1}^{N}g_{s,n}g_{D,n}\right]\right)\nonumber \\
 & =\log_{2}\left(1+\frac{P_{s}r_{s}^{-\beta}r_{D}^{-\beta}}{N_{0}}\mathbb{E}\left[\sum_{n=1}^{N}g_{s,n}g_{D,n}\right]\right),\label{eq:cd-aprx-1-1}
\end{align}
where $\gamma_{D}^{r}$ is defined in (\ref{eq:snr-D-r}), as the
instantaneous SNR at $D$ for the VANET RIS relay network. Given that
the channels $g_{s}$ and $g_{D}$ are independent, then the expectation
in (\ref{eq:cd-aprx-1-1}) can be obtained as 
\begin{align}
\mathbb{E}\left[\sum_{n=1}^{N}g_{s,n}g_{D,n}\right] & =\sum_{n=1}^{N}\mathbb{E}\left[g_{s,n}\right]\mathbb{E}\left[g_{D,n}\right]\nonumber \\
 & \stackrel{\left(d\right)}{=}\stackrel[0]{\infty}{\int}\stackrel[0]{\infty}{\int}\sum_{n=1}^{N}g_{s}^{2}g_{D}^{2}K_{0}\left(g_{D}\right)e^{-\frac{g_{s}^{2}}{2}}\textrm{d}g_{D}\textrm{d}g_{s}\nonumber \\
 & \stackrel{\left(e\right)}{=}N\left(\frac{\pi}{2}\right)^{\frac{3}{2}},\label{eq:mean-trpl-ray}
\end{align}
where $(d)$ in (\ref{eq:mean-trpl-ray}) was obtained using (\ref{eq:meijer-bessel})
and the fact that $g_{s}$ is Rayleigh distributed with PDF $f\left(g_{s}\right)=g_{s}\exp\left(\frac{1}{2}g_{s}^{2}\right)$,
while $(e)$ was obtained with the aid of \cite[Eqs. (6.521.10) and (3.326.2)]{book2}.
From (\ref{eq:cd-aprx-1-1}) and (\ref{eq:mean-trpl-ray}), we obtain
a bounded expression for the average capacity at $D$ for the VANET
RIS relay as
\begin{equation}
\overline{C}_{D,r}^{aprx}=\log_{2}\left(1+\frac{NP_{s}\left(\pi\right)^{\frac{3}{2}}r_{s}^{-\beta}r_{D}^{-\beta}}{2\sqrt{2}N_{0}}\right).\label{eq:cd-aprx-final-r}
\end{equation}

Using similar analysis to the derivation of (\ref{eq:cd-aprx-final-r}),
we obtain a bounded expression for the average capacity at $E$ for
the VANET RIS relay as
\begin{equation}
\overline{C}_{E,r}^{aprx}=\log_{2}\left(1+\frac{NP_{s}\left(\pi\right)^{\frac{3}{2}}r_{s}^{-\beta}r_{E}^{-\beta}}{2\sqrt{2}N_{0}}\right).\label{eq:ce-aprx-final-r}
\end{equation}

From (\ref{eq:cs-defined-2}), (\ref{eq:cd-aprx-final-r}) and (\ref{eq:ce-aprx-final-r}),
we obtained the desired bounded expression for the VANET RIS relay
as 
\begin{equation}
\overline{C}_{s,r}^{aprx}=\log_{2}\left(\frac{2\sqrt{2}N_{0}+NP_{s}\left(\pi\right)^{\frac{3}{2}}r_{s}^{-\beta}r_{D}^{-\beta}}{2\sqrt{2}N_{0}+NP_{s}\left(\pi\right)^{\frac{3}{2}}r_{s}^{-\beta}r_{E}^{-\beta}}\right).\label{eq:cs-aprx-final-r}
\end{equation}

The accuracy of (\ref{eq:cs-aprx-final-r}) as compared to the exact
analytical solution in (\ref{eq:Cs-final-r}) will be demonstrated
in the discussion in Sec. \ref{sec:Results}.

\subsection{Secrecy Outage Probability}

In this section, we derive an expression for the SOP of the VANET
RIS relay system. Similar to the derivation in Sec. \ref{subsec:SOP-1},
from (\ref{eq:cs-defined-2}) and (\ref{eq:pout-defined}) we obtain

\begin{align}
P_{o}^{r} & =\textrm{Pr}\left[\log_{2}\left(\frac{1+\gamma_{D}}{1+\gamma_{E}}\right)<c_{\textrm{th}}\right]\nonumber \\
 & =\textrm{Pr}\left[\frac{1+\gamma_{D}}{1+\gamma_{E}}<2^{c_{\textrm{th}}}\right]\nonumber \\
 & \stackrel{\left(f\right)}{=}\textrm{Pr}\left[P_{s}r_{s}^{-\beta}r_{D}^{-\beta}\sum_{n=1}^{N}g_{s,n}g_{D,n}<\right.\nonumber \\
 & \hphantom{}\qquad\quad\left.\nu\left(N_{0}+P_{s}r_{s}^{-\beta}r_{E}^{-\beta}\sum_{n=1}^{N}g_{s,n}g_{E,n}\right)-N_{0}\right]\nonumber \\
 & \stackrel{\left(g\right)}{=}\textrm{Pr}\left[\sum_{n=1}^{N}g_{s,n}g_{D,n}<\underset{\varTheta_{r}}{\underbrace{\frac{N_{0}\left(\nu-1\right)}{P_{s}r_{s}^{-\beta}r_{D}^{-\beta}}+\frac{\nu r_{E}^{-\beta}}{r_{D}^{-\beta}}\sum_{n=1}^{N}g_{s,n}g_{E,n}}},\right]\label{eq:pout-defined-3}
\end{align}
where $\nu=2^{c_{\textrm{th}}}$ and $\left(f\right)$ follows from
(\ref{eq:snr-D-r}) and (\ref{eq:snr-E-r}). Owing to the difficulty
in obtaining a direct expression for the distribution of the sum $\sum_{n=1}^{N}g_{s,n}g_{D,n}$
in $\left(g\right)$ of (\ref{eq:pout-defined-3}), we employ the
CLT to approximate with a Gaussian distribution with parameters; mean
$\mu_{r}=N\left(\frac{\pi}{2}\right)^{\frac{3}{2}}$ (from (\ref{eq:mean-trpl-ray}))
and variance $\sigma_{r}^{2}$ given by
\begin{align}
\textrm{Var}\left[\sum_{n=1}^{N}g_{s,n}g_{D,n}\right] & =\mathbb{E}\left[\sum_{n=1}^{N}g_{s,n}^{2}g_{D,n}^{2}\right]-\mathbb{E}\left[\sum_{n=1}^{N}g_{s,n}g_{D,n}\right]^{2}\nonumber \\
 & =N\left(8-\left(\frac{\pi}{2}\right)^{\frac{3}{2}}\right).\label{eq:var-of-sum-1}
\end{align}

Thus from (\ref{eq:pout-defined-3}), the SOP can be given by 
\begin{equation}
P_{o}\approx\frac{1}{2}\left(1+\textrm{erf}\left(\frac{\varTheta_{r}-\mu_{r}}{\sqrt{2}\sigma_{r}}\right)\right),\label{eq:sop-aprx-defined-1}
\end{equation}
where $\varTheta_{r}$ is defined in (\ref{eq:pout-defined-3}) and
$\mathbb{E}\left[\varTheta_{r}\right]$ can be computed as
\begin{align}
\mathbb{E}\left[\varTheta_{r}\right] & =\mathbb{E}\left[\frac{N_{0}\left(\nu-1\right)}{P_{s}r_{s}^{-\beta}r_{D}^{-\beta}}+\frac{\nu r_{E}^{-\beta}}{r_{D}^{-\beta}}\sum_{n=1}^{N}g_{s,n}g_{E,n}\right]\nonumber \\
 & =\mathbb{E}\left[\frac{N_{0}\left(\nu-1\right)}{P_{s}r_{s}^{-\beta}r_{D}^{-\beta}}+\frac{\nu r_{E}^{-\beta}}{r_{D}^{-\beta}}\sum_{n=1}^{N}g_{s,n}g_{E,n}\right]\nonumber \\
 & \stackrel{\left(h\right)}{=}\frac{N_{0}\left(\nu-1\right)}{P_{s}r_{s}^{-\beta}r_{D}^{-\beta}}+\frac{\nu N\pi^{3}r_{E}^{-\beta}}{2\sqrt{2}r_{D}^{-\beta}},\label{eq:mean-theta-r}
\end{align}
where $\left(h\right)$ in (\ref{eq:mean-theta-r}) follows from (\ref{eq:mean-trpl-ray})
with $\mathbb{E}\left[\sum_{n=1}^{N}g_{s,n}g_{E,n}\right]=\frac{N\pi^{3}}{2\sqrt{2}}.$
The SOP for the VANET RIS relay system is therefore
\begin{equation}
P_{o}^{r}\approx\frac{1}{2}\left(1+\textrm{erf}\left(\frac{\frac{N_{0}\left(\nu-1\right)}{P_{s}r_{s}^{-\beta}r_{D}^{-\beta}}+\frac{N\pi^{3}}{2\sqrt{2}}\left(\frac{\nu r_{E}^{-\beta}}{r_{D}^{-\beta}}-1\right)}{\sqrt{2N\left(8-\left(\frac{\pi}{2}\right)^{\frac{3}{2}}\right)}}\right)\right).\label{eq:sop-aprx-final-r}
\end{equation}

\section{\textcolor{black}{Numerical Results and Discussions\label{sec:Results}}}

\textcolor{black}{In this section, we present and discuss some results
from the mathematical expressions derived in the paper. We then investigate
the effects of key parameters on the secrecy capacity of the system.
The results are then verified using Monte Carlo simulations with at
least $10^{4}$ iterations. Unless otherwise stated, we have assumed
}source power $P_{s}=10$W, RIS-to-$D$ distance $r_{D}=4$m, RIS-to-$E$
distance $r_{E}=8$m, source-to-RIS distance $r_{s}=10$m, secrecy
outage threshold $c_{\textrm{th}}=1$ bit/Hz and path loss exponent
$\beta=2.7$.

\begin{figure}[th]
\begin{centering}
\includegraphics[scale=0.45]{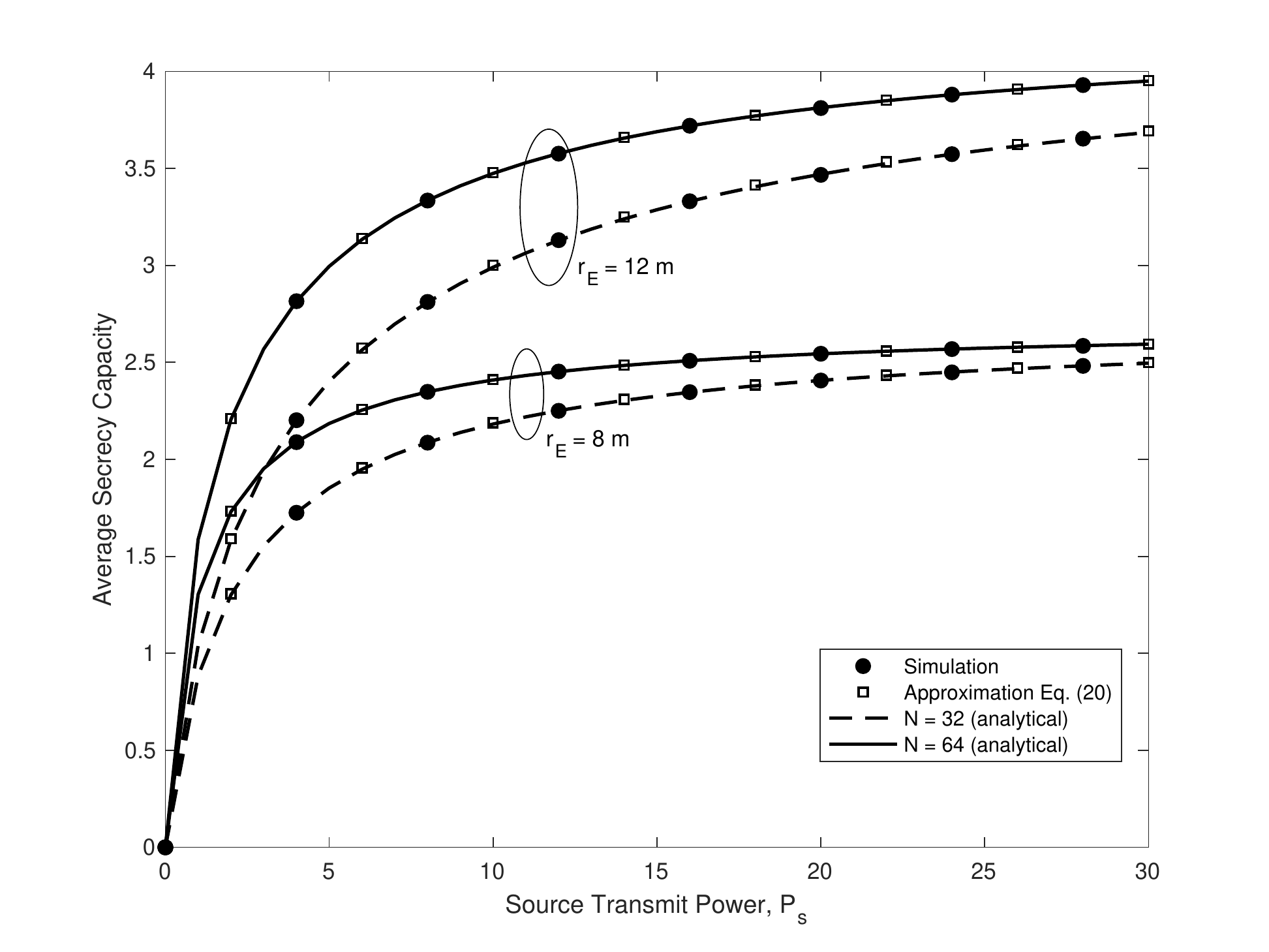}
\par\end{centering}
\caption{Average secrecy capacity versus source transmit power $P_{s}$ for
the V2V network with RIS as AP. Parameters considered with varying
eavesdropper distance $r_{E}$ and number of RIS cells $N$. \label{fig:cs-vs-ps-v2v}}
\end{figure}
\textcolor{black}{}
\begin{figure}[th]
\begin{centering}
\textcolor{black}{\includegraphics[scale=0.45]{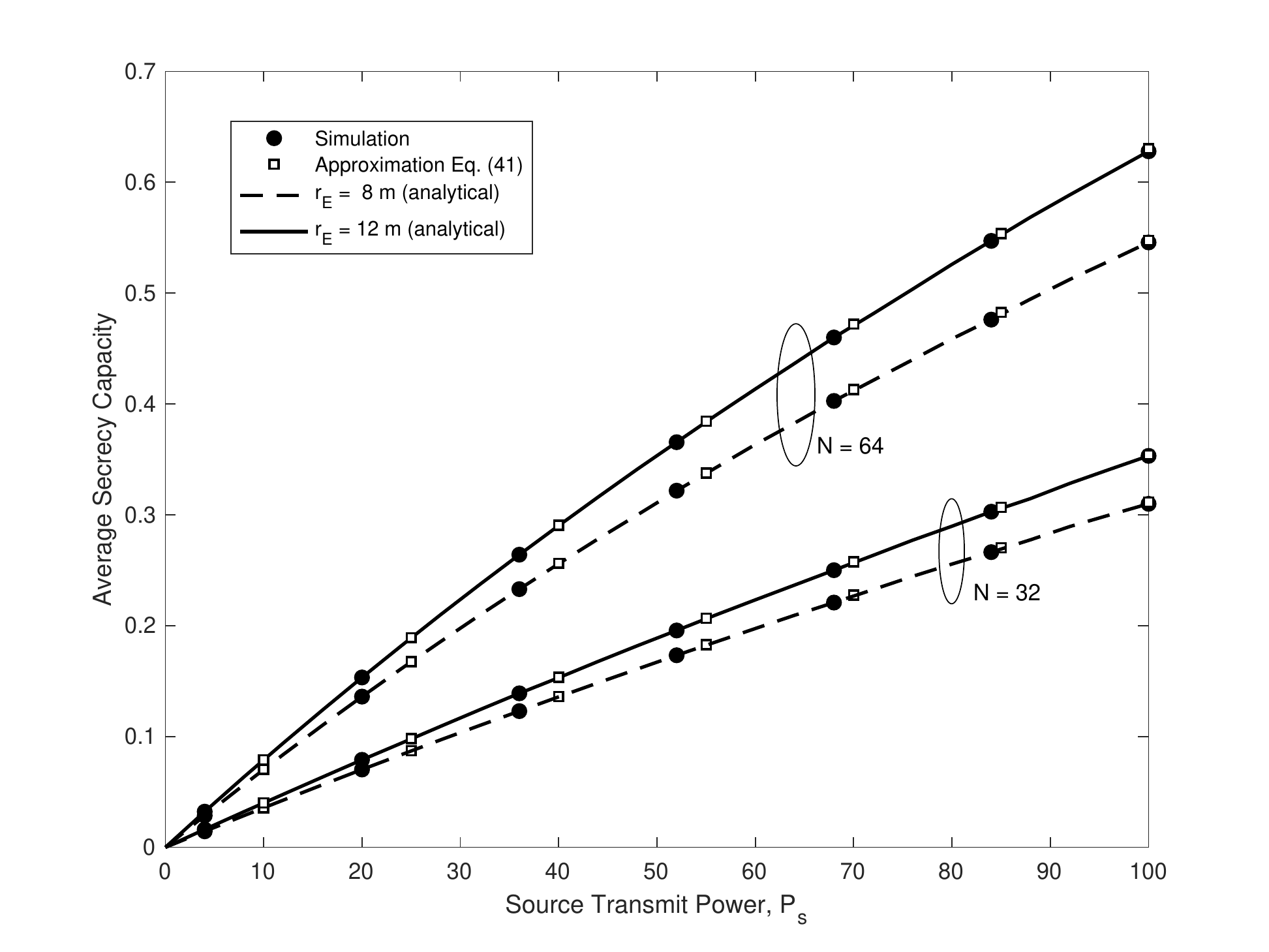}}
\par\end{centering}
\textcolor{black}{\caption{Average secrecy capacity versus source transmit power $P_{s}$ for
the VANET with RIS as relay. Parameters considered with varying eavesdropper
distance $r_{E}$ and number of RIS cells $N$.\label{fig:cs-vs-ps-relay}}
}
\end{figure}
\textcolor{black}{}
\begin{figure}[th]
\begin{centering}
\textcolor{black}{\includegraphics[scale=0.45]{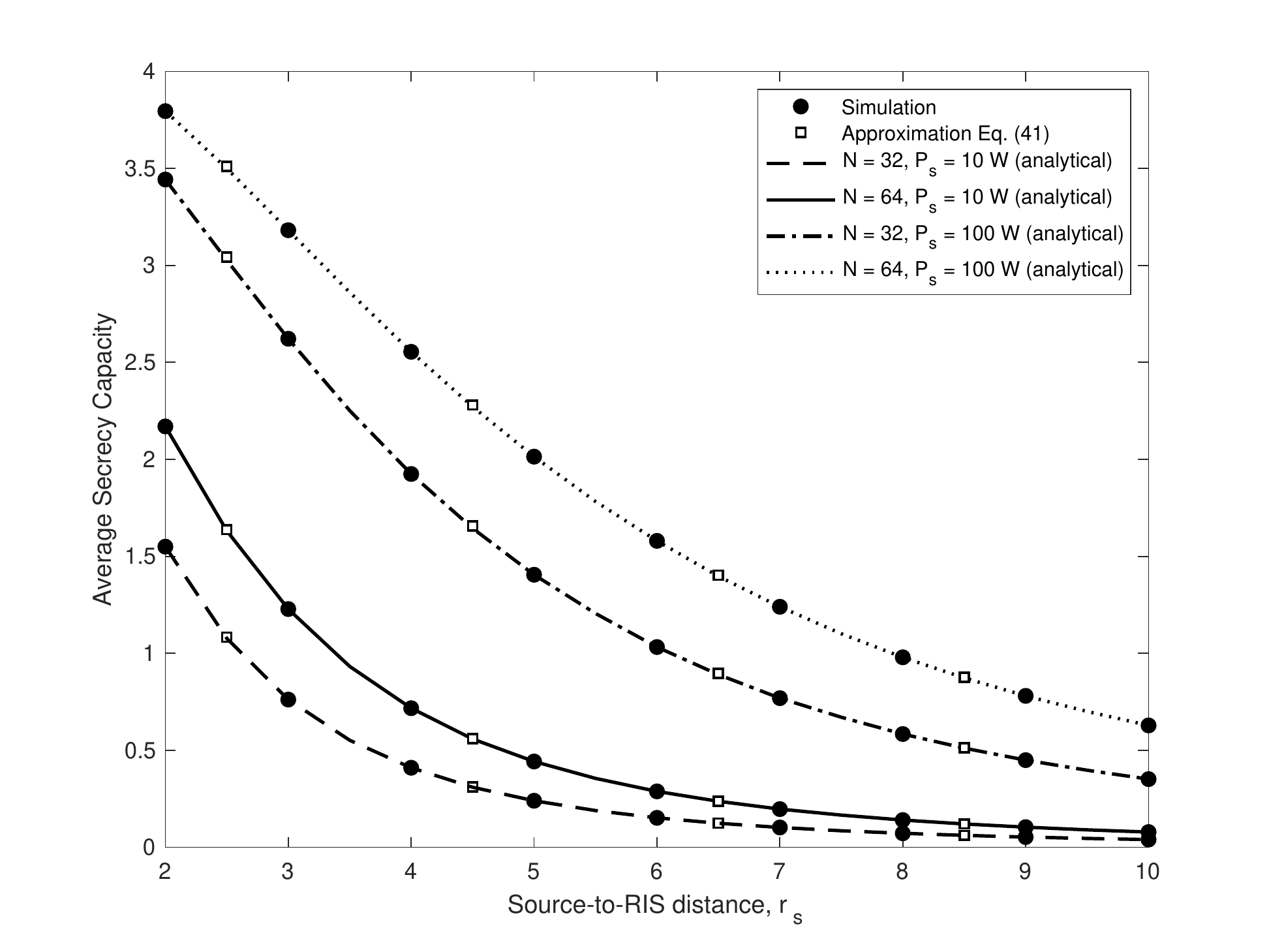}}
\par\end{centering}
\textcolor{black}{\caption{Average secrecy capacity versus source-to-RIS distance for the VANET
with RIS as relay. Parameters considered with varying source transmit
power $P_{s}$ and number of RIS cells $N$.\label{fig:cs-vs-rs-relay}}
}
\end{figure}
\textcolor{black}{}
\begin{figure}[th]
\begin{centering}
\textcolor{black}{\includegraphics[scale=0.45]{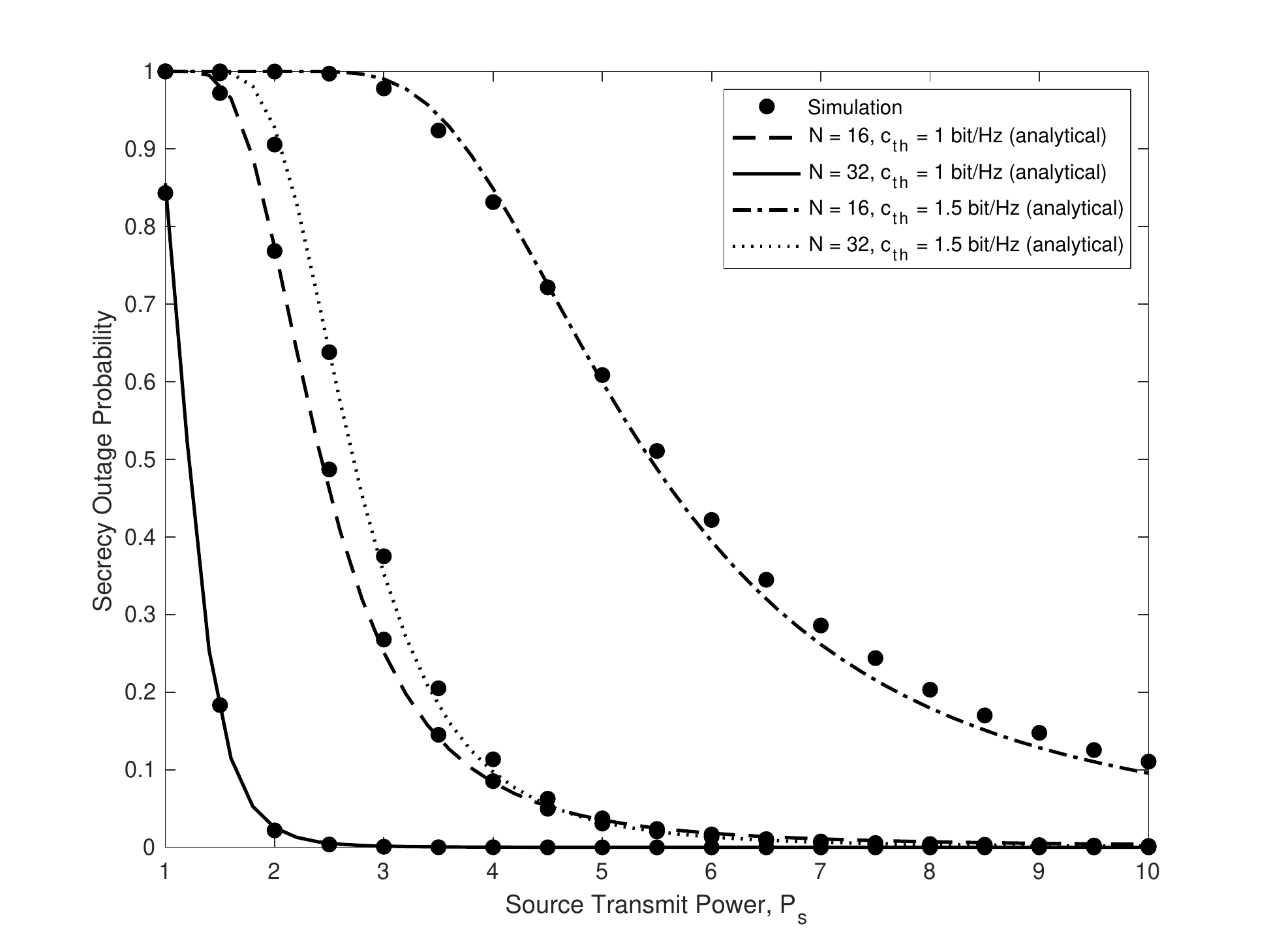}}
\par\end{centering}
\textcolor{black}{\caption{Secrecy outage probability versus source transmit power $P_{s}$ for
the V2V network with RIS as AP. Parameters considered with varying
secrecy capacity threshold $c_{\textrm{th}}$ and number of RIS cells
$N$.\label{fig:sop-vs-ps-v2v}}
}
\end{figure}

\subsection{Average Secrecy Capacity}

In this subsection, we examine the ASC of both system models considered.
In Fig. \ref{fig:cs-vs-ps-v2v}, we commence analysis for the V2V
network model, with the ASC against source power for different numbers
of RIS cells and eavesdropper distances. It can be observed that the
secrecy capacity increases with an increase in $P_{s}$, $r_{E}$
or $N$. It can be further noted that within the region considered,
the eavesdropper distance has a greater effect on the secrecy capacity
than doubling the number of RIS cells. Also, the effect of increased
RIS cells, is more pronounced when the eavesdropper is further away.
A similar analysis can be made for the ASC of the VANET RIS-relay
model considered, as observed in Fig. \ref{fig:cs-vs-ps-relay}. However,
the effect of increased source power produces an almost linear response
for the secrecy capacity, while the effective value of the secrecy
capacity is much lower than the V2V RIS model for similar $P_{s}$.
In both Figs. \ref{fig:cs-vs-ps-v2v} and \ref{fig:cs-vs-ps-relay},
it can also be observed that the analytical results in (\ref{eq:cs-aprx-final})
and (\ref{eq:cs-aprx-final-r}) provide highly accurate approximations
for the secrecy capacity of the V2V RIS model and VANET RIS-relay
model respectively.

Fig. \ref{fig:cs-vs-rs-relay}, shows a plot of the ASC against $r_{s}$
with different values of $P_{s}$ and $N$, for the VANET RIS-relay
system. We assume the RIS-to-eavesdropper distance to be $r_{E}=12$m.
First, we observe that the ASC decreases as the source distance increases,
demonstrating the effect of fading and path loss on the link, before
the RIS relay. 

The result also demonstrates that doubling the number of RIS cells
has less influence on the ASC, as compared to the impact of the source
power, within the observed region. Furthermore, we again observe the
high accuracy of the approximate expressions in (\ref{eq:cs-aprx-final-r}),
within the regions of interest studied in these results.

\subsection{Average Secrecy Outage Probability}

In this subsection, we present results for the average SOP of both
systems considered from the expressions derived. In Fig. \ref{fig:sop-vs-ps-v2v},
the SOP of the V2V RIS access point is plotted against the source
transmit power for different threshold values and numbers of RIS cells.
From the plot, we observe that the SOP decreases with increased transmit
power or increase number of RIS cells, resulting from stronger SNR
conditions. On the other hand, the SOP is proportional to the threshold
within the region investigated, as expected. We can particularly observe
that the SOP performance is affected more significantly when the threshold
is at 1.5 bit/Hz compared to 1 bit/Hz, such that over half an order
of magnitude gain in performance is possible, when the RIS cells are
doubled from $N=16$ to 32, in some regions of operation for this
system model. \textcolor{black}{}
\begin{figure}[th]
\begin{centering}
\textcolor{black}{\includegraphics[scale=0.45]{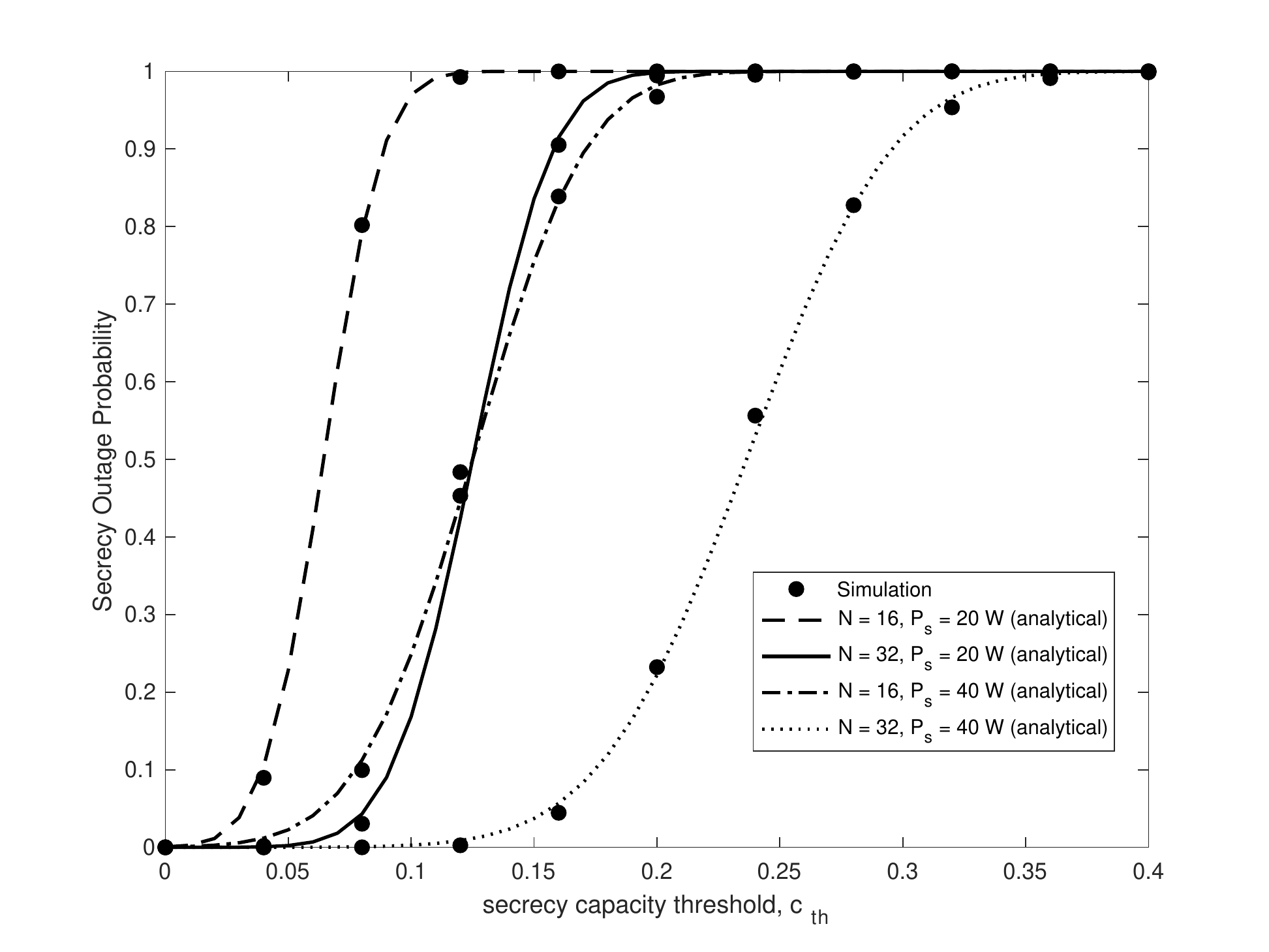}}
\par\end{centering}
\textcolor{black}{\caption{Secrecy outage probability versus secrecy capacity threshold $c_{\textrm{th}}$
for the VANET with RIS as relay. Parameters considered with varying
source transmit power $P_{s}$ and number of RIS cells $N$.\label{fig:sop-vs-cth-relay}}
}
\end{figure}

\begin{figure}[th]
\begin{centering}
\textcolor{black}{\includegraphics[scale=0.45]{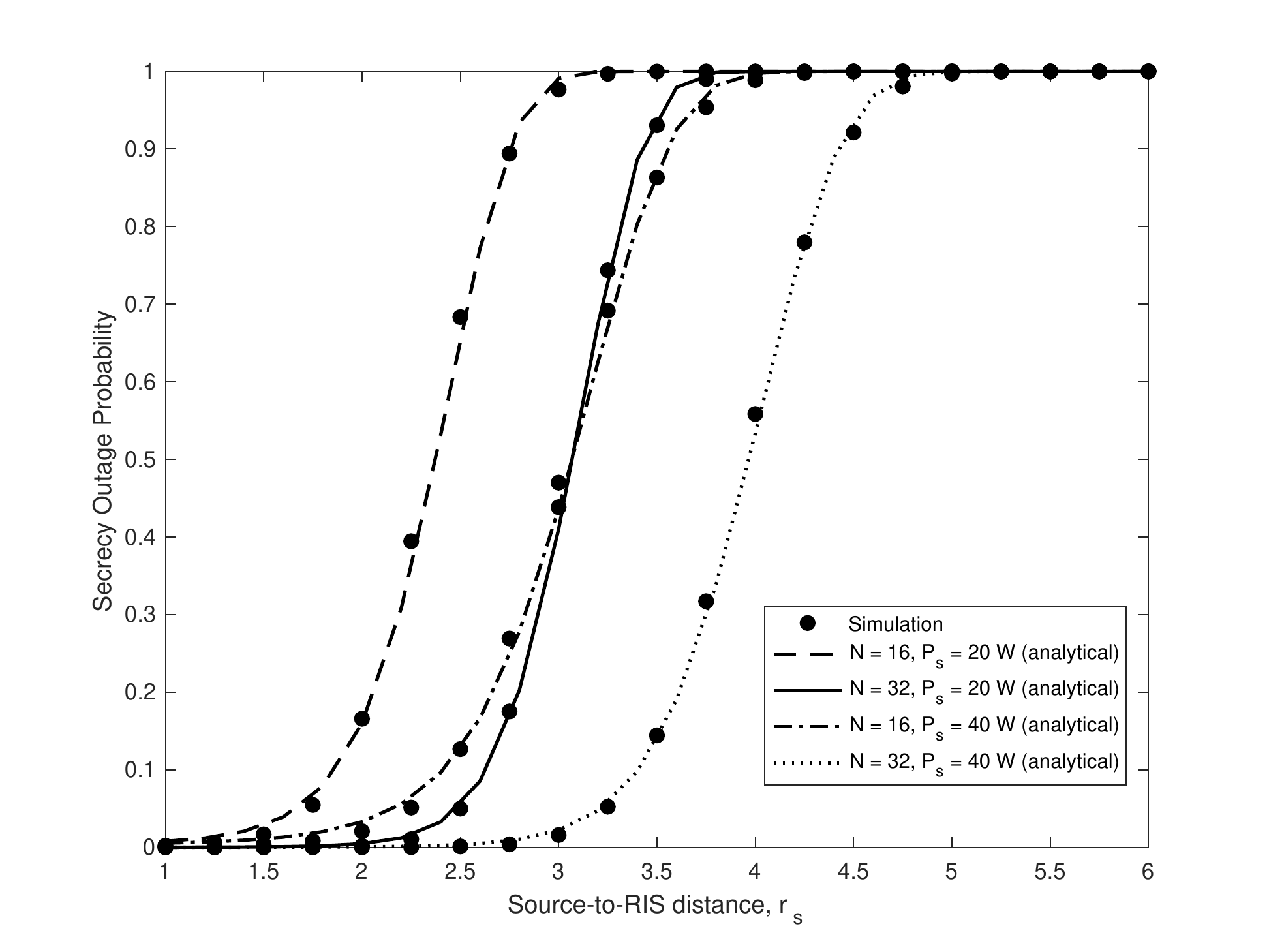}}
\par\end{centering}
\textcolor{black}{\caption{Secrecy outage probability versus source-to-RIS distance for the VANET
with RIS as relay. Parameters considered with varying source transmit
power $P_{s}$ and number of RIS cells $N$.\label{fig:sop-vs-rs-relay}}
}
\end{figure}

The effect of the system parameters on the SOP for the VANET with
RIS relay system is illustrated in Figs. \ref{fig:sop-vs-cth-relay}
and \ref{fig:sop-vs-rs-relay}. In Fig. \ref{fig:sop-vs-cth-relay},
the SOP for the VANET with RIS is plotted against the expected secrecy
threshold, for different values of source transmit power and number
of RIS cells. It can be observed that for this model, the system performance
is degraded (increased SOP) when we adopt higher threshold values
for all corresponding values of $N$ and $P_{s}$. It is interesting
to see that, the effect of doubling the number of cells from $N=16$
to 32 at a fixed transmit power is more pronounced than doubling the
power from $P_{s}=20$ to 40 W, with respect to the SOP. In Fig. \ref{fig:sop-vs-rs-relay},
the SOP for the VANET with RIS is plotted against the source-to-RIS
distance, where the SOP curves show decreased performance when $r_{s}$
increases and that the SOP rapidly degrades within a short distance.
For example, it can be shown that within the regions observed, the
SOP falls by about an order of magnitude (from SOP = 1 to 0.1) for
a change of less than 2 m. Moreover, we observe in this case, the
effect of changing $N=16$ to 32 at a fixed transmit power is similar
to increasing the power from $P_{s}=20$ to 40 W, with respect to
the SOP. 

\textcolor{black}{}

Therefore, from a practical perspective, the results in Figs. \ref{fig:sop-vs-cth-relay}
and \ref{fig:sop-vs-rs-relay} indicate that there is a design decision
to consider the cost versus benefit of extra hardware required to
double the RIS cells or the source transmit power, while noting that
for some applications the number of RIS cells could be much larger
than 100. (i.e. $N>>100$).

\section{\textcolor{black}{Conclusions\label{sec:Conclusions}}}

\textcolor{black}{In this paper, we examined the effects of key parameters}
on the PLS of a wireless vehicular communication network. As a novel
study on the subject, two scenarios of a RIS-based vehicular network
were considered. The results demonstrate how the secrecy capacity
and SOP of a vehicular network can be improved with respect to the
source power, eavesdropper distance, the number of RIS cells, the
source-to-relay distance and the secrecy threshold. The results further
showed how the location and size of RIS (in terms of number of RIS
cells) can be employed to improve a RIS relay-based VANET, while clearly
indicating the benefits of employing the RIS in all cases. 

\bibliographystyle{IEEEtran}
\bibliography{bibGC19}

\end{document}